\documentclass[pdflatex,sn-mathphys-num]{sn-jnl}

\usepackage{graphicx}%
\usepackage{multirow}%
\usepackage{amsmath,amssymb,amsfonts}%
\usepackage{amsthm}%
\usepackage{mathrsfs}%
\usepackage[title]{appendix}%
\usepackage{xcolor}%
\usepackage{textcomp}%
\usepackage{manyfoot}%
\usepackage{booktabs}%
\usepackage{algorithm}%
\usepackage{algorithmicx}%
\usepackage{algpseudocode}%
\usepackage{listings}%
\usepackage{siunitx}

\theoremstyle{thmstyleone}%
\theoremstyle{thmstyletwo}%

\theoremstyle{thmstylethree}%

\begin{document}

\title[Brilliance: toward autonomous computational research in particle accelerator physics]{Brilliance: toward autonomous computational research in particle accelerator physics}


\author*[1,2]{\fnm{Jinyu} \sur{Wan}}\email{wanjinyu@ihep.ac.cn}

\author[1,3]{\fnm{Yu} \sur{Zhao}}\email{zhaoyu@ihep.ac.cn}

\author[1,2]{\fnm{Wei} \sur{Bao}}\email{baowei@ihep.ac.cn}

\author[1,2]{\fnm{Ying} \sur{Li}}\email{liy02@ihep.ac.cn}

\author[1,2]{\fnm{Hao} \sur{Zhong}}\email{zhonghao@ihep.ac.cn}

\author[1,3]{\fnm{Xiaohan} \sur{Lu}}\email{luxh@ihep.ac.cn}

\author*[1,2]{\fnm{Yi} \sur{Jiao}}\email{jiaoyi@ihep.ac.cn}

\affil[1]{\orgdiv{Institute of High Energy Physics}, \orgname{Chinese Academy of Sciences}, \orgaddress{\city{Beijing}, \postcode{100049}, \country{China}}}

\affil[2]{\orgname{University of Chinese Academy of Sciences}, \orgaddress{\city{Beijing}, \postcode{100049}, \country{China}}}

\affil[3]{\orgname{Spallation Neutron Source Science Center}, \orgaddress{\city{Dongguan}, \postcode{523803}, \country{China}}}


\abstract{
Computational research requires repeated decisions about models, calculations and the interpretation of results. We introduce Brilliance, an agent that combines large language models, reusable physics expertise and numerical simulation to construct and revise investigations in accelerator physics. Starting from a detailed expert-specified request, Brilliance develops an ideal storage-ring lattice meeting the specified emittance ceiling and sampled nonlinear-acceptance criteria after revising optics that fail particle-tracking tests. Further scientist-guided exploration produces a candidate with a natural emittance of 6.51 pm$\cdot$rad and millimeter-scale transverse acceptance, while exposing unresolved momentum-acceptance limitations. Using measured responses from the High Energy Photon Source, Brilliance also performs offline optics calibration consistent with an independent specialist analysis. An expert-module ablation examines how reusable domain knowledge influences research choices and outcomes. Here, we demonstrate a practical route to scientist-agent collaboration in sustained numerical research.
}




\maketitle

\section{Introduction}\label{sec1}

Numerical simulation is central to research across condensed-matter physics, plasma physics, astrophysics, nuclear and particle physics, and accelerator physics, bridging theoretical models to complex phenomena and proposed experiments \cite{foulkes2001qmc,dawson1983plasma,vogelsberger2014galaxies,Carlson2015quantum,durr2008hadron,shiltsev2021colliders}. Its scientific value depends not only on numerical accuracy but also on how an investigation is formulated and revised. Researchers are required to choose appropriate approximations, construct informative calculations and determine what an unexpected result implies for the next step. Translating these decisions into software remains labor-intensive and can restrict the range of hypotheses and designs explored. This creates an opportunity for automation at the level of the investigation, supporting the repeated interaction between scientific interpretation and numerical experiments, rather than accelerating individual calculations alone.

Large language models (LLMs) offer a means of connecting scientific knowledge with executable actions. Over the past few years, Coscientist and SciToolAgent established tool-mediated experimental automation and multi-tool workflow orchestration \cite{boiko2023coscientist,ding2025scitool}, while FunSearch and Empirical Research Assistance coupled generated programs to numerical evaluation and search, extending automation from tool invocation to program construction and empirical optimization \cite{romera2024funsearch,aygun2026era}. The Virtual Lab and Robin advanced this trajectory by distributing scientific reasoning across collaborating agents, enabling experimental design and autonomous hypothesis generation \cite{swanson2025virtuallab,ghareeb2026robin}. More recently, MutexaGPT bridged qualitative physical intuition and quantitative simulation by translating natural-language design intent into physics-based molecular modeling and variant proposals \cite{shao2026mutexa}, and AI-driven approaches have begun to propose entirely new experimental layouts in physics, moving beyond parameter tuning \cite{klimesch2026designing}. These advances establish program construction and scientific revision as substantive directions of agent research. They also motivate examining how domain knowledge influences the calculations an agent chooses to perform, and whether those choices improve scientific outcomes.

In particle accelerator field, machine learning-based methods have already accelerated multiobjective design, stabilized synchrotron source properties and optimized coupled laser-plasma parameters \cite{leemann2019stability,edelen2020surrogate,shalloo2020bayesian}. Language-model agents have subsequently enabled accelerator tuning from natural-language instructions and coordinated multistage experiments involving data retrieval, script generation, machine interaction and analysis \cite{kaiser2025tuning,hellert2026agentic}. These methods, however, remain largely confined to tuning parameters or invoking tools within a predefined computational framework. The open question is whether an agent can consistently maintain a scientific objective across successive calculations while revising its approach to pursuing that objective. 

The coupled requirements and interpretive challenges of accelerator physics provide a stringent testbed for this capability. The demanding physics requirements of accelerators are tightly coupled with practical engineering constraints. For instance, in modern storage-ring accelerator design, bending and focusing are coordinated to reduce beam emittance, while nonlinear magnets are used to maintain chromatic correction and stable particle motion over the required amplitude and momentum ranges \cite{raimondi2021hmba,riemann2019lgb}. The investigations can require different numerical strategies, with the physical question guiding the selection and interpretation of calculations. The resulting conclusions are then evaluated against scientific criteria, a process that relies heavily on expert judgment.

Here we introduce Brilliance (\textbf{B}eam \textbf{R}esearch \textbf{I}ntelligence with \textbf{L}arge \textbf{L}anguage-model \textbf{I}ntegration for \textbf{A}ccelerator \textbf{N}umerical \textbf{C}omputation and \textbf{E}xploration), an LLM-based research agent that combines reusable physics knowledge with task-specific programs executed through the JuTrack simulation engine \cite{wan2025jutrack}. Starting from a scientist's objectives, Brilliance constructs numerical studies, interprets their outputs and revises subsequent calculations within a research project (Fig. 1). The agent is guided by expert modules that encapsulate reusable domain knowledge, including diagnostic criteria, numerical methods, and physical constraints specific to accelerator physics. The system also maintains a dedicated workspace that retains programs, models, and results, allowing earlier candidates and interpretations to be revisited. Conventional algorithms optimize parameters within a fixed numerical study, whereas Brilliance can redefine the variables, constraints, or diagnostics that define the next study. This separation makes it possible to adapt how an investigation is formulated without replacing the underlying physics algorithms.

\begin{figure}[h]
\centering
\includegraphics[width=0.8\textwidth]{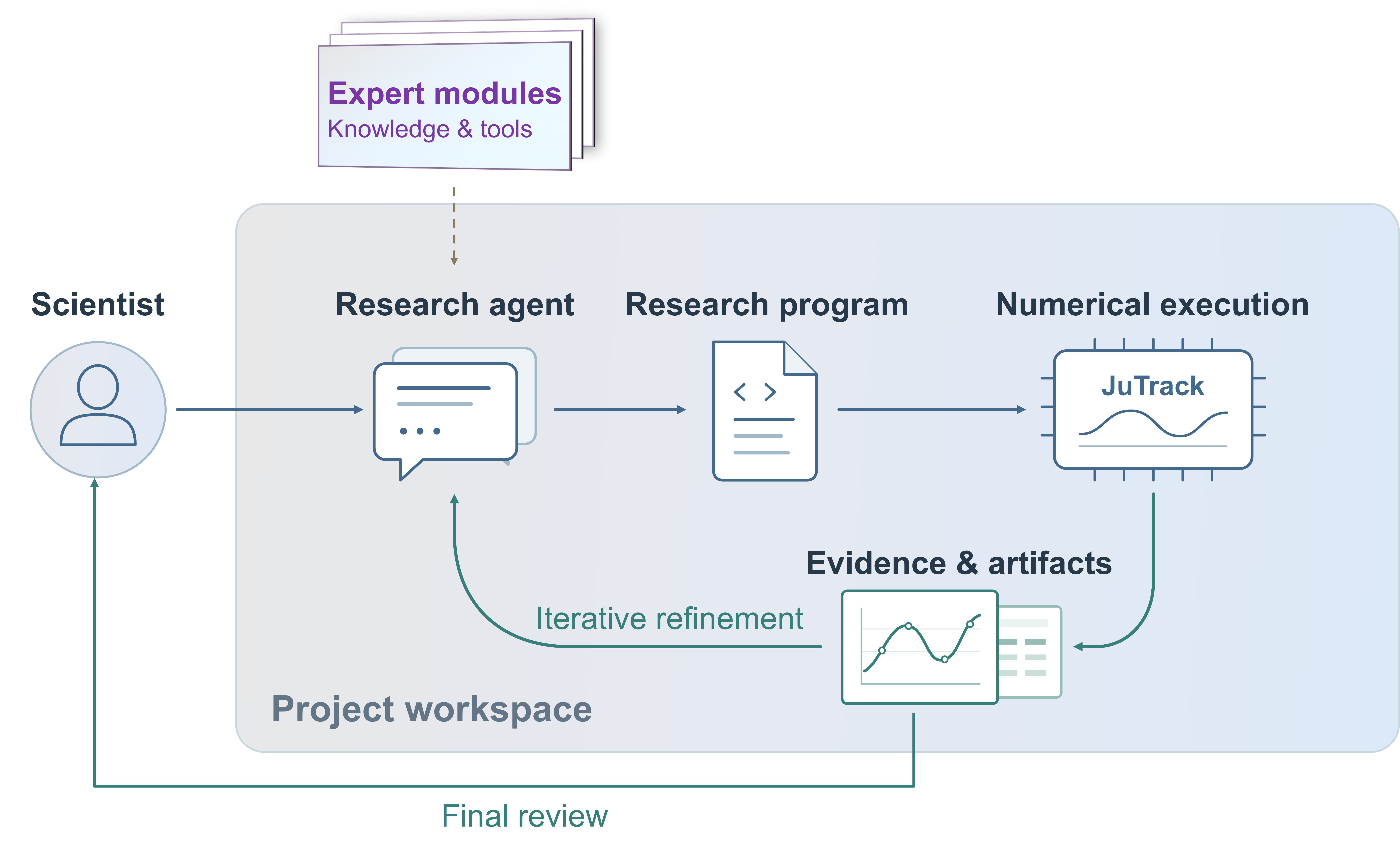}
\caption{Overview of the Brilliance research workflow. A scientist specifies the research objectives and constraints. Guided by expert modules that provide reusable physics knowledge and computational tools, the large language model-based research agent constructs task-specific programs for numerical execution through JuTrack. The agent interprets the resulting evidence and artifacts to revise subsequent calculations, forming an iterative cycle of program construction, execution and scientific interpretation. The project workspace retains programs, models and results, allowing earlier candidates and calculations to be revisited. The resulting evidence and final outputs are returned to the scientist for review}\label{fig1}
\end{figure}

We evaluate Brilliance on tasks of designing new synchrotron light sources and calibrating the model of an operating light source. The first task produces an ideal lattice satisfying coupled emittance and acceptance requirements. Further exploration with scientist guidance yields a sub-10-pm superperiodic lattice with sufficiently large dynamic aperture under practical engineering constraints. Using measured orbit-response and dispersion data from the High Energy Photon Source (HEPS) \cite{jiao2018heps}, Brilliance calibrates machine optics consistent with an independent expert analysis. Expert-module ablation and a comparison between language models examine how guidance and model choice affect progress on the design task. Together, these studies demonstrate the extent to which an agent can sustain numerical investigation through repeated construction, interpretation and revision of calculations.

\section{Results}\label{sec2}
\subsection{Multistage design of a diffraction-limited synchrotron light source}\label{sec:results-1}
The design of a synchrotron light source aims to produce bright, spatially coherent X-rays from a stored electron beam in a storage ring. Its central challenge is to define the magnetic lattice including the arrangement, lengths, and field strengths of the magnets that guide and focus the electrons. The lattice sets the natural emittance, i.e., the equilibrium transverse phase-space area of the beam. Reducing natural emittance enhances both the attainable photon brilliance and the coherent fraction \cite{raimondi2023extremely}. A viable lattice must preserve particles over a sufficient range of transverse amplitudes and momentum deviations. These properties, characterized by the dynamic aperture (DA) and momentum acceptance (MA), govern injection efficiency and beam lifetime, and must be achieved under realistic engineering constraints within a finite circumference while reserving straight sections for radiation-producing insertion devices and space for accelerator equipment.

We asked Brilliance to construct the magnetic lattice of a synchrotron light source from scratch with only one chat of about 120 words,
\begin{quote}
\small
``Start a new 3.5\,GeV, 810\,m, 32-cell modified-H7BA design.
The first goal is a validated baseline near
$20\,\mathrm{pm}\cdot\mathrm{rad}$, accepting $\leq 30\,\mathrm{pm}$,
with chromaticities $(+5,+5)$, useful connected transverse acceptance
and local momentum acceptance larger than $\pm2\%$.
Reserve a clear 6\,m achromatic straight per cell and account for
services separately within the circumference.
Use outer longitudinal-gradient bends with embedded focusing in B1/B7,
and B3-B5 modules with higher-field centers, defocusing combined-function
flanks and neighboring reverse bends.
Include independent defocusing quadrupoles between B3/B4 and B4/B5,
at least three sextupole sites in each bump, and up to two octupole
sites per cell.
Parameterize internal lengths, bend angles, field profiles and gradients
and do not freeze the first geometry.''
\end{quote}
The chat message contains the basic design constraints of electron beam energy and storage-ring circumference, as well as the objective combined low natural emittance with a sufficiently large acceptance for on-momentum and off-energy particles. A specific request of using dispersion bumps is raised for the lattice layout design, which would force the agent to perform a hybrid multi-bending-achromatic (MBA) lattice design instead of a standard MBA that is less practical or even unlikely for the extremely high design target. It is worth noting that, to further benchmark Brilliance's design capability against that of experienced accelerator physicists, we intended to adopt the same parameter specifications as those used in the latest Southern Advanced Photon Source (SAPS) \cite{ZHAO2026171310} design, a planned medium-energy fourth-generation light source in southern China.

\begin{figure}[h]
\centering
\includegraphics[width=0.8\textwidth]{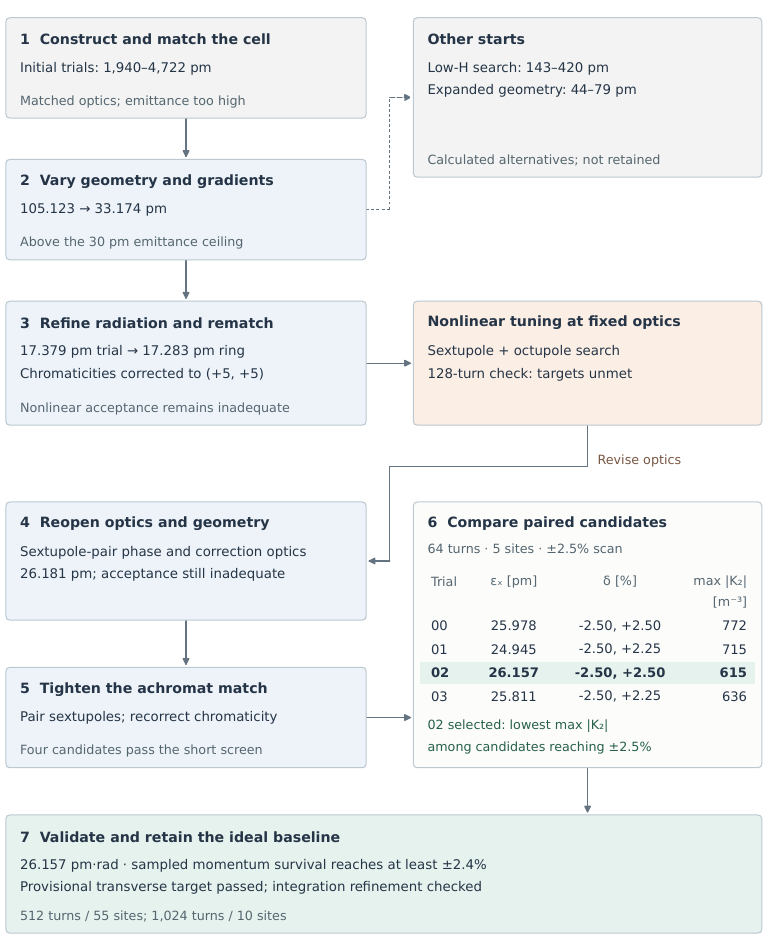}
\caption{Workflow of Brilliance for designing a 3.5\,GeV synchrotron light source targeting a natural emittance below \(30\,\mathrm{pm}\cdot\mathrm{rad}\). The agent starts with a linear optics design and proceeds to nonlinear dynamics checks. When these checks fail, it changes strategy to rebuild the linear optics, and ultimately meets the nonlinear requirements through nonlinear beam dynamics optimization.}\label{fig2}
\end{figure}

The investigation process of Brilliance is summarized in Fig. 2 and the original data of the full process is available at Ref. \cite{jinyu2026example}. Within this scope, Brilliance is responsible for constructing the parameterization, developing the simulation and optimization programs, and selecting the calculations needed to evaluated and improve the candidates. The initial radiation and chromatic-correction studies reaches an apparently promising solution near $17.283\,\mathrm{pm}\cdot\mathrm{rad}$ that has met the preferred target. However, chromatic correction and short particle-tracking screens exposes poor nonlinear acceptance. Low emittance and corrected chromaticities are therefore insufficient reasons to retain the optical solution. Brilliance then changes the formulation of the search, reopening the linear-optics and geometry search while incorporating the phase relationships of sextupole pairs and chromaticity correction performance into the considerations. It generates several re-matched alternatives and progressively tightens the achromat constraints before evaluating the complete periodic cells. Particle-tracking simulation then distinguishes candidates that looked similar by their radiation properties. 

\begin{figure}[h]
\centering
\includegraphics[width=0.9\textwidth]{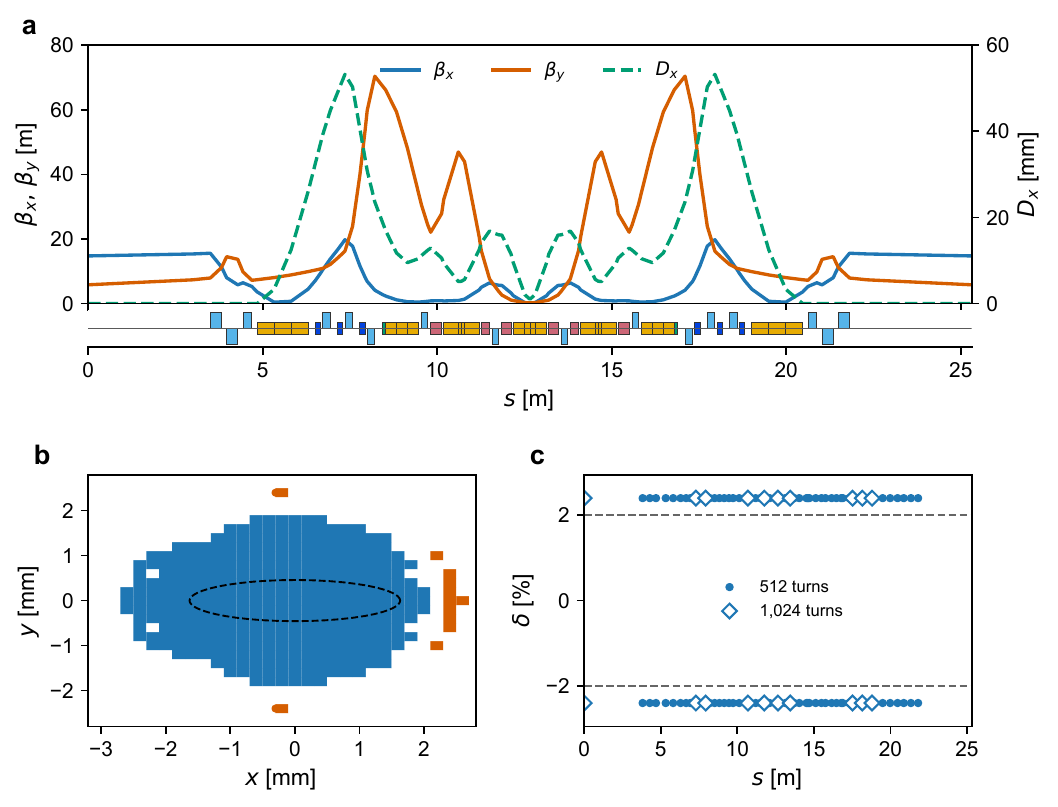}
\caption{A single periodic cell of the final design proposed by Brilliance for the 3.5\,GeV diffraction-limited synchrotron light source design task.
\textbf{a}, Lattice layout and betatron oscillation optics. Yellow, light blue, deep blue, and pink blocks denote bending, quadrupole, sextupole, and reverse-bending magnets, respectively.
\textbf{b}, DA area in $x-y$ space of the lattice obtained from 1,024-turn particle tracking. The black circle indicates the $3\sigma$ range, where $\sigma$ is the beam size. The blue region indicates the connected surviving region for particles, whereas the orange regions indicate disconnected surviving regions.
\textbf{c}, Local MA of the lattice. Round and diamond markers represent the local MA evaluated using 512-turn and 1,024-turn particle tracking, respectively.}\label{fig3}
\end{figure}

For the ultimately selected optics shown in Fig.~\ref{fig3}(a), the worst sampled local MA of 512-turn full ring tracking expands from [-1.25\%, +1.00\%] to at least [-2.40\%, +2.40\%] through nonlinear beam dynamics optimization, with emittance unchanged at approximately $26.16\,\mathrm{pm}\cdot\mathrm{rad}$ and chromaticities of $(+5,+5)$. Here the $\pm 2.40\%$ values are scan-limited lower bounds, not measured loss boundaries. Each periodic cell preserves the 6-m clear straight and a separate 1-m service allocation within the 810-m circumference. Since the synchrotron radiation damping for an electron storage ring usually takes over a few thousand of turns, to avoid insufficient particle tracking, Brilliance performs another 1,024-turn tracking simulation with the same sampled momentum offsets at 10 selected longitudinal locations to ensure that the particle could still survive (Fig.~\ref{fig3}(b)). 

Transverse stability is another important concern in the design of modern storage-ring accelerators. Fig.~\ref{fig3}(c) shows that after 1,024 turns, the DA boundary reaches nearly 2 mm in both horizontal and vertical planes. Brilliance is targeting the DA area with matched injected geometric emittance of 20 and 4 $\mathrm{nm}\cdot\mathrm{rad}$, corresponding to $3\sigma$ position half-axes of 1.630 and 0.458 mm, which are study assumptions rather than a scientist-specified injection distribution. 

Traditional diffraction-limited storage-ring design typically requires years of specialist effort and thousands of CPU cores on high-performance computing clusters. In contrast, this campaign demonstrates that the agent establishes an ideal-lattice baseline that satisfies the emittance ceiling and nonlinear acceptance requirements within four hours on a personal computer, requiring only a single chat interaction with the user. The agent sacrifices a lower-emittance optics solution in favor of one with better chromatic correction and nonlinear acceptance. Such a trade-off is not trivial to identify manually, as it requires balancing competing objectives through extensive iterative studies.

\subsection{Toward sub-10-pm regime}\label{sec:results-2}
For hard X-rays at a wavelength of \(0.1\,\mathrm{nm}\), the diffraction-limited emittance is approximately \(8\,\mathrm{pm}\cdot\mathrm{rad}\). Reducing the emittance to the 1 pm level would significantly enhance brightness, bring the transverse coherence of soft X-rays close to unity, and markedly improve that of hard X-rays, thereby enabling novel coherent scattering and imaging experiments. Recent studies indicate that a sub-10-pm lattice can be realized with a compact superperiodic structure \cite{zhaoipac2026tup2612}. However, optimizing the nonlinear dynamics of such a compact focusing structure remains extremely challenging, and only impractically limited DA and MA have been reported.

To further explore the \(1\,\mathrm{pm}\)-level design, a new task is assigned to Brilliance with the chat message,
\begin{quote}
\small
``Design from scratch a 3.5 GeV electron storage ring below \(10\,\mathrm{pm}\cdot\mathrm{rad}\), within 1 km circumference, with at least 36 achromatic insertion straights averaging at least 4 m. Explore a jointly matched two-arc superperiod with many short bends interleaved with quadrupoles, one localized dispersion bump per arc, and an intermediate user straight. Optimize the layout as well as individual bend lengths, angles and strengths; retain quadrupole focusing inside the bumps to create complementary sextupole-site optics, consider longitudinal-gradient and reverse bends, and jointly improve emittance, chromatic correction and nonlinear acceptance within practical engineering limits.''
\end{quote}
The beam-energy requirement remains the same as in Sect.~2.1, while the circumference is relaxed to \(1{,}000\,\mathrm{m}\) and the specific targets for chromaticity and nonlinear acceptance are not directly mentioned. In addition to the design target, we briefly add several approaches that may enable a \(1\,\mathrm{pm}\)-level design. 

\begin{figure}[h]
\centering
\includegraphics[width=0.9\textwidth]{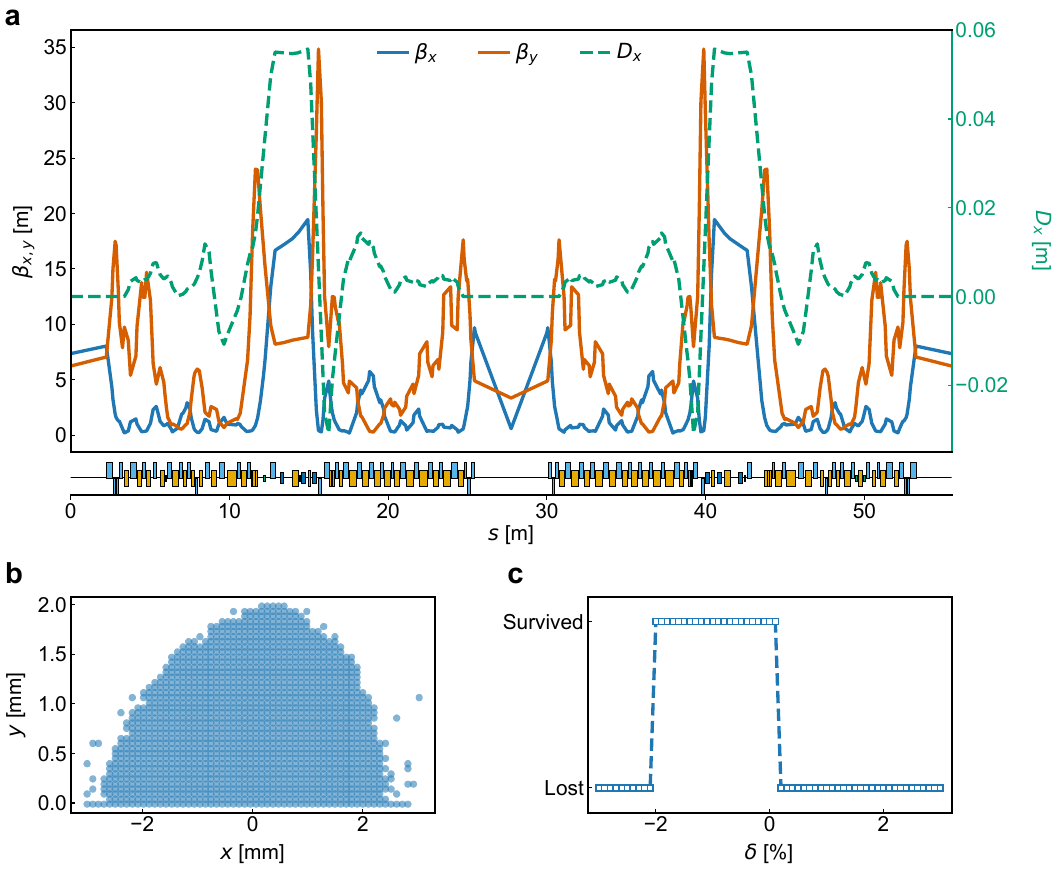}
\caption{A single periodic cell of the \(1\,\mathrm{pm}\)-level lattice proposed by Brilliance.
\textbf{a}, Lattice layout and optics. The element color definition is consistent with Fig. 3.
\textbf{b}, DA area in $x-y$ space of the obtained lattice. The blue points represent particles that could survive over 1,024-turn particle tracking.
\textbf{c}, MA of the lattice at the center of the straight section. The horizontal axis represents the scanned momentum-deviation range, and the vertical axis indicates whether the particle survives or is lost.}\label{fig4}
\end{figure}

Similar to the design process in Sect.~2.1, Brilliance first achieves a natural emittance of \(4.514\,\mathrm{pm}\cdot\mathrm{rad}\) within the compact geometry under the survey constraints (the full design process is available in Ref.~\cite{jinyu2026example}). However, particle tracking reveals only a small transverse DA, and redistributing the sextupole strengths yields little improvement. Brilliance then utilizes the emittance margin below \(10\,\mathrm{pm}\cdot\mathrm{rad}\) to explore modifications to the underlying optics, shifting its objective from minimizing emittance to recovering transverse stability within the target regime. Nevertheless, no valid solutions with \(1\,\mathrm{pm}\)-level emittance and acceptable DA are found with Brilliance, and it reports reaching its stopping criterion.

New scientist-guided chat messages are provided to Brilliance to optimize the octupoles and the fluctuation of resonance-driving terms in the lattice to further improve the acceptance. After the fourth conversation, Brilliance produces several candidate lattices. We select the one having largest DA with a natural emittance of \(6.51\,\mathrm{pm}\cdot\mathrm{rad}\). It has 18 superperiods and 36 achromatic 4.5-m straight sections within a \(998.927\,\mathrm{m}\) circumference (Fig.~4). Its layout differs substantially from that of the proposed \(5.57\,\mathrm{pm}\cdot\mathrm{rad}\) design in Ref.~\cite{zhaoipac2026tup2612}, which features systematic dispersion suppression in the optics. As shown in Fig.~4(b), its on-momentum DA reaches nearly 3\,mm horizontally and 2\,mm vertically, approximately three times the values reported in Ref.~\cite{zhaoipac2026tup2612} in each plane.

However, the MA remains challenging to improve. Fig. 4(c) shows that the selected candidate has a MA of [-2.0\%, 0.1\%] at the center of the straight section, making it infeasible for a real light source. Even after providing a new prompt for free exploration of the MA, the issue persisted. Solving this problem may require further adjustment of the linear optics to improve the chromaticity-correction efficiency and reduce the nonlinearity from the correction sextupoles. Nevertheless, this exploration, driven by the iterative scientist-agent interaction, advances a compact superperiodic architecture toward the \(1\,\mathrm{pm}\) level with millimetre-scale DA, demonstrating the potential of human-agent collaboration to push the design beyond conventional limits.

\subsection{Reconstructing an operating accelerator from measurements}\label{sec:results-3}
In a real particle accelerator facilities, magnet field errors, misalignment, and calibration uncertainties break the designed periodicity, causing $\beta$-beating and coupling that degrade brightness, coherence, and injection efficiency of the facility. A widely-used approach to calibrate these errors is LOCO (Linear Optics from Closed Orbits) \cite{safranek1997loco}, which iteratively fits the measured responses of the machine to a lattice model and adjusts magnet strengths to restore the designed optics. However, setting up and evaluating a machine-specific LOCO analysis requires expert judgment to select fitting parameters, interpret residual patterns, and decide which corrections to apply. The process demands substantial specialist experience and repeated human-machine interaction, making it a natural test case for evaluating whether an autonomous agent can navigate the same iterative diagnosis-and-correction loop with minimal human input.

We evaluate Brilliance using measured data from the HEPS \cite{jiao2018heps}, a 6 GeV fourth-generation synchrotron light source located in Beijing, China, which began operation in late 2025. The HEPS storage ring is equipped with 577 beam position monitors (BPMs) and 384 fast correctors, which can be used to measure orbit-correction responses. The ring also contains 528 independently powered quadrupoles and several specially designed sextupole movers. Both the strengths of the quadrupoles and the positions of the movers require calibration. In practice, magnetic errors are mixed with BPM and corrector calibration errors, making the calibration problem coupled. 

Since three BPMs are unavailable, calibration is performed using a \(574 \times 384\) response matrix and the ring dispersion measured under operating conditions. Brilliance is then tasked with calibrating the lattice model using these measurements. We use only one chat message to interact with Brilliance by indicating the location of the data and lattice file, the paired variables that must be tuned together, and BPM gains need to calibrate. For comparison, an independent LOCO calibration is also performed by experienced experts.

Brilliance begins by verifying the correspondence between the measured channels and lattice elements, after which a coupled fitting program is constructed in JuTrack to jointly fit the response matrix and dispersion. BPM calibrations are adjusted alongside the magnetic parameters, so that discrepancies in the measured response can be attributed to both lattice and instrument errors. The final refinement converges after 39 full iterations, and the response calculation is subsequently verified using finite difference check. The resulting lattice and calibration parameters are then saved without further scientific instructions.

\begin{figure}[t]
    \centering
    \includegraphics[width=\textwidth]{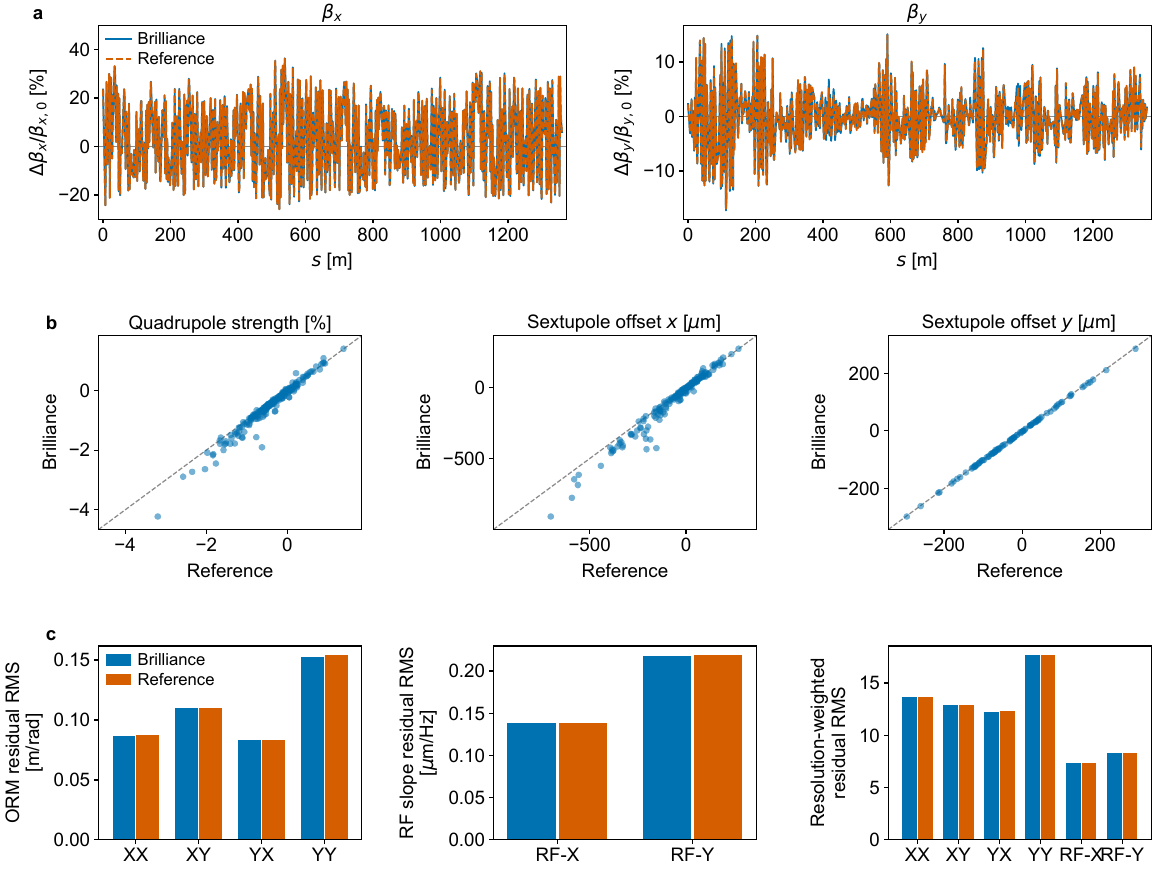}
    \caption{Comparison of Brilliance and human-expert calibration results for HEPS using measured orbit-response and dispersion data.
    \textbf{a}, Horizontal and vertical $\beta$-beating inferred by Brilliance (solid blue) and an independent, physicist-written LOCO implementation (dashed orange).
    \textbf{b}, Comparison of fitted quadrupole-strength deviations and horizontal and vertical sextupole mover position offsets. Each point represents a corresponding fitted parameter. The dashed lines indicate equality.
    \textbf{c}, RMS residuals of the four orbit-response blocks (XX, XY, YX and YY), horizontal and vertical dispersion, and the corresponding BPM-resolution-normalized residual RMS.}
    \label{fig:5}
\end{figure}

The reconstructed optics closely reproduces the expert reference throughout the ring (Fig.~\ref{fig:5}a). Both analyses identify the same longitudinal pattern of horizontal and vertical $\beta$-beating, including the substantial local departures from the design optics. The consistency observed across the two calibration tasks raises concerns of optical distortions caused by machine errors, rather than from fitting errors.

The fitted magnetic parameters show corresponding agreement (Fig.~\ref{fig:5}b). Most quadrupole-strength deviations and sextupole offsets lay close to the identity line, with particularly tight agreement for the vertical sextupole offsets. Slightly larger differences occur for a subset of quadrupole strengths and horizontal sextupole offsets. Nevertheless, the reference comparison supports consistency of the reconstructed machine optics without establishing unique recovery of all underlying errors.

The residual comparison evaluates how well each model represents the measurements (Fig.~\ref{fig:5}c). Brilliance and the expert reference obtain nearly identical RMS residuals across all four orbit-response blocks and the dispersion, in physical units and after normalization by the supplied BPM-resolution scales. Both analyses retain discrepancies larger than the adopted resolution scales. Their consistency therefore establishes comparable fitting performance rather than reproduction of the measurements at the BPM-resolution level. These results demonstrate that Brilliance can autonomously perform an offline inverse analysis and produce a machine-specific calibration that closely matches the specialist reference from a single task specification, with no additional manual intervention.

\subsection{Expert-module ablation and model comparison}\label{sec:results-4}
To examine how reusable domain guidance and the underlying language model influence agent performance, we compare additional Brilliance runs against the baseline presented in Sect.~\ref{sec:results-1}. That baseline uses GPT-6 Astra-High with plug-in expert modules of MBA design. In the first comparison, we retain the same model but omit the optional expert modules. The second comparison retains the expert modules but replace the model with DeepSeek-4.1 Flash. All runs start from the same design request and share the core execution module, JuTrack API guides, research tools, and web access. 

Table~\ref{tab:1} summarizes the comparison. Without the expert modules, Brilliance constructs an explicit 810-m ring meeting the natural emittance and chromaticity requirements but fails the requested acceptance. The retained candidate has a MA of [-1.25\%,+0.75\%] at the straight-section center and the agent reports the design as incomplete. In contrast, the expert-guided run successively rematches sextupole-pair phase relationships and correction optics, producing candidates that pass subsequent tracking. The ablated run produces nearly three times as many language-model responses as the expert-guided baseline. Further investigation shows that it performs twelve web searches, compared with only four online material readings in the baseline, indicating greater reliance on external information when expert modules are absent.

With DeepSeek-4.1 Flash and expert modules enabled, the run does not produce a design satisfying the same request. Intermediate calculations yield an achromatic lattice near $679.84\,\mathrm{pm}\cdot\mathrm{rad}$ with chromaticities close to (+5,+5), exceeding the emittance ceiling. Subsequent attempts encounter infeasible optimization bounds and fail to establish a compliant low-emittance solution. The run reaches its research budget without validated DA or MA. DeepSeek-4.1-Flash produces 394 language-model responses and consumes 27.57M tokens, significantly more than the baseline. The higher token usage reflects more frequent tool calls and repeated unsuccessful attempts, indicating a more extensive and less efficient exploration process rather than a larger number of design alternatives. Nevertheless, the cost of different models varies considerably, with DeepSeek-4.1 Flash being approximately 60 times cheaper than GPT-6 Astra. Although it fails the complicated light-source design task, cheaper models remain an economical option for simpler tasks.

\begin{table}[t]
\centering
\caption{Comparison of Brilliance runs under different configurations. One language-model response may contain multiple tool calls. The longer wall time in the Baseline case mainly arises from the nonlinear beam dynamics optimization, which requires massive particle tracking simulations. In contrast, the other two cases are primarily limited by linear optics optimization, which is much faster.}\label{tab:1}
\small
\setlength{\tabcolsep}{4pt}
\renewcommand{\arraystretch}{1.18}
\begin{tabular}{@{}p{0.23\textwidth}p{0.24\textwidth}p{0.24\textwidth}p{0.23\textwidth}@{}}
\hline
Configuration & Baseline & Ablation & Model comparison \\
\hline
Language models & GPT-6 Astra-High & GPT-6 Astra-High & DeepSeek-4.1 Flash \\
Expert modules & Enabled & Disabled & Enabled \\
Horizontal emittance ($\mathrm{pm}\cdot\mathrm{rad}$) & 26.157 & 29.500 & 679.84 \\
Chromaticities & $(+5,+5)$ & $(+5,+5)$ &  $(+5,+5)$ \\
MA &  $[-2.4\%,+2.4\%]$ & $[-1.25\%,+0.75\%]$ & Not established \\
Model response steps & 64 & 184 & 394 \\
Elapsed wall time (h) & 3.21 & 1.79 & 5.66 \\
Token usage (M) & 8.25 & 6.98 & 27.57 \\
\hline
\end{tabular}
\end{table}

\section{Discussion}\label{sec12}
This study demonstrates how a scientific agent can sustain an investigation in which the numerical formulation evolves as evidence accumulates. In the storage-ring design, a low-emittance solution fails the nonlinear acceptance checks, prompting the agent to reconsider the linear optics and geometry in order to ultimately satisfy the coupled requirements. The significance of this progression lies in using failed validation to revisit how the problem is posed. Together with the HEPS reconstruction, it shows that program generation can support scientific work that requires repeated decisions about what to calculate and how to interpret the results.

The expert-module ablation suggests that recognizing a physical difficulty and resolving it require different capabilities. Without the modules, the agent identifies excessive sextupole demand and revises its search, yet does not recover sufficient MA. In contrast, the guided run translates related design principles into a successful sequence of rematching and tracking studies. Conversely, supplying expert modules to a different model did not ensure a compliant design. These observations suggest that the value of reusable expertise depends on the agent's ability to translate it into effective numerical choices while maintaining the task constraints. 

The more ambitious sub-10-pm exploration highlights the role of scientist-agent collaboration. Guidance from scientists extends the search beyond the agent's initial stopping point and yields millimetre-scale DA, about three times the previously reported values, although inadequate momentum acceptance still prevents a viable light-source design. This outcome exposes a remaining physical obstacle and provides a basis for further investigation. Autonomy in this context therefore refers to sustained work within a specified scientific scope. The initial requests already contain substantial architectural knowledge, and the more difficult task requires additional expert intervention. 

Brilliance provides a practical route to delegate substantial parts of numerical investigation while retaining scientific judgment over objectives and validation. By coupling reusable expert knowledge with explicit assumption tests, records of unsuccessful searches, and criteria for requesting additional guidance, it makes the rationale for each computational decision easier to examine and allows scientists to focus on the questions that remain unresolved. More broadly, this approach makes complex numerical studies easier to construct, revise, and scrutinize, thereby expanding the range of investigations that researchers can pursue. Although currently tailored to accelerator physics, Brilliance can be extended to other numerical domains by equipping it with appropriate simulation tools and domain guidance.

\section{Methods}\label{sec:methods}

\subsection{Brilliance framework}
Brilliance is implemented in Python and integrates the open-source Codex agent runtime \cite{codex2025} for conversation management and iterative tool use, while supporting native tool-calling loops. Brilliance supports multiple compatible language-model backend, such as the OpenAI API and the DeepSeek API. A local model context protocol (MCP) server is implemented for writing and revising research programs.

For each investigation, Brilliance supplies the agent with the user request, relevant expert knowledge, numerical interface descriptions and the available project history. The agent translates the request into executable calculations, choosing the study-specific parameterization, objectives, constraints and diagnostic quantities. It can revise these choices as numerical evidence accumulates. This allows an investigation to develop through candidate construction, evaluation and further exploration, with the calculation sequence determined by the research question and intermediate results. The scientist can inspect the resulting programs and evidence, refine the objectives or redirect the investigation through subsequent requests.

We implement a built-in Python service layer that connects agent-written research programs to JuTrack \cite{wan2025jutrack} and provides reusable operations for accelerator studies. A common lattice representation supports construction, modification, and storage of accelerator models, with an adapter that translates element definitions and physical parameters into native JuTrack objects and checks the supported element parameters. The \texttt{accelerator\_api} provides a common interface for lattice construction and modification, optics and radiation calculations, chromaticity evaluation, particle tracking, and parameterized lattice matching. Its evaluation routines associate numerical results with the evaluated candidate and distinguish periodic-cell calculations from explicit full-ring evaluations.

Several specialized services support nonlinear and optimization studies. \texttt{acceptance\_measurements} provides means of calculating DA and MA for storage-ring accelerators. \texttt{response\_analysis} constructs response matrices for specified observables with respect to tuning knobs, with diagnostics of matrix rank and conditioning. \texttt{population\_optimization} provides population-based optimization tool for single or multiple objectives. \texttt{parallel\_evaluation} distributes independent calculations across separate worker processes on multi-CPU machines, each with its own numerical environment and mutable lattice.

The agent combines these built-in services with Python numerical libraries to construct study-specific programs, selecting the variables, objectives, constraints, and diagnostic calculations. Native JuTrack operations remain accessible when a study requires capabilities beyond the provided services. 

For complex tasks, existing guide and tools can be insufficient. \texttt{web\_search} allows the agent to search and read online materials as a supplement. Additional accelerator expertise is organized into independently loadable modules centered on an \texttt{EXPERT.md} document. Each module describes the applicable physical regimes, methods, constraints, diagnostic criteria and supporting references for a specific domain. This knowledge guides both the formulation of calculations and the interpretation of their results. New modules can be added without modifying the core agent controller. Taking the synchrotron light source design task presented in Sect~\ref{sec2} as an example, an MBA-design module is installed in Brilliance. It provides guidance on the relationships among radiation properties, lattice optics, sextupole correction, and nonlinear acceptance. 

To support investigations spanning multiple calculations and model interactions, each project is assigned a dedicated workspace. This workspace retains the conversation history, program source, inputs, accelerator lattices, and numerical outputs. Results are linked to the calculation that produced them and, where applicable, to the evaluated lattice candidate, so that later work can compare alternatives or continue from a selected configuration. At each turn boundary, the agent indicates whether to continue, conclude the request, or seek scientist input, subject to a configured continuation budget. The framework records whether the research objective has been satisfied separately from execution completion, allowing an investigation to conclude with a partial result or an unresolved limitation.

\subsection{Lattice design and particle tracking}
The storage-ring design tasks in Sect.~\ref{sec2} start from prescribed beam energy, geometric constraints and performance targets. Brilliance constructs the lattice parameterizations and search programs within these requirements using JuTrack, which supplies the optics, chromaticities, radiation integrals and particle trajectories. Natural emittance of a storage-ring accelerator is the equilibrium state resulting from the balance between synchrotron radiation damping and quantum excitation. It is calculated from the radiation integrals as
\begin{equation}
\epsilon_{x,0}
= \frac{C_q \gamma^2}{J_x} \frac{I_5}{I_2},
\end{equation}

\begin{equation}
I_5 = \oint \frac{\mathcal{H}}{|\rho|^3}\,\mathrm{d}s,
\qquad
I_2 = \oint \frac{1}{\rho^2}\,\mathrm{d}s,
\end{equation}

\begin{equation}
    \mathcal{H} = \gamma_x\eta^2 + 2\alpha_x\eta\eta' + \beta_x\eta'^2\
\end{equation}
where \(C_q = 55/(32\sqrt{3})\,\hbar/(m_e c) \approx 3.8319\times 10^{-13}\,\mathrm{m}\), \(\gamma\) is the Lorentz factor, \(J_x\) is the horizontal damping partition number, $\alpha_x$, $\beta_x$, and $\gamma_x$ are the Courant-Snyder parameters, and $\eta$ is the dispersion.

Besides natural emittance, DA and MA are also important considerations in modern storage-ring accelerator design. DA is the range of transverse amplitudes over which particles remain stable over many turns, setting the transverse acceptance limit of the lattice, whereas MA is the maximum relative momentum deviation that particles can tolerate while remaining stable, determining the longitudinal acceptance and strongly influencing beam lifetime. Both are evaluated by tracking finite samples of initial conditions and retaining the stable region connected to the closed orbit. Typically, a lepton ring requires at least 1,000 turns of particle tracking to determine DA and MA, while shorter tracking can serve as a faster indicator but still requires validation by long-term tracking.

\subsection{Machine-model calibration}
The calibration task in Sect.~\ref{sec:results-3} uses a coupled LOCO fit \cite{safranek1997loco} of measured orbit-response matrices and dispersion data to the supplied $6\,\mathrm{GeV}$ HEPS lattice. The response matrix is measured at 574 BPM locations with 384 corrector channels. The dispersion is measured by scanning the radio-frequency (RF) of the cavities around \(166.6\,\mathrm{MHz}\) over a total span of \(10\,\mathrm{Hz}\). Across 48 periodic cells, the fitting includes five quadrupole-strength groups and six groups of sextupole mover offset per cell, giving 528 magnetic parameters. BPM gains and cross-plane calibration terms and corrector gains bring the total to 3,204 fitted parameters.

Model responses and dispersions are calculated from JuTrack. For the \(k\)-th active observation, the normalized residual is
\begin{equation}
r_k(\boldsymbol{\theta})
= \frac{F_k(\boldsymbol{\theta}) - y_k}{s_k},
\end{equation}
where \(\boldsymbol{\theta}\) is the vector of calibration parameters, \(F_k(\boldsymbol{\theta})\) is the model calculation for observation \(k\) under \(\boldsymbol{\theta}\), \(y_k\) is the corresponding measurement, and \(s_k\) is the supplied resolution scale.

The fitting objective is
\begin{equation}
 \Phi(\boldsymbol{\theta})=\sum_{k\in\mathcal{M}}w_k^2r_k^2,
\end{equation}
where $\mathcal{M}$ contains the $N$ observations. The additional residual weight \(w_k\) is set to 20 for dispersion observations and to 1 otherwise, and this weighting is applied in both the Brilliance and specialist calibrations. 

In the specialist calibration, we use the MATLAB LOCO code \cite{safranek2002matlab}, which has been successfully applied to model calibration and optics correction at many synchrotron light sources. The program calculates the Jacobian of the observations with respect to the fitted parameters and adjusts the fitting parameters in the lattice model to minimize the deviation $\Phi(\boldsymbol{\theta})$ between the model prediction and the measurements using the Gauss-Newton method. The Jacobian is calculated through finite-difference approximation in the model.

For the Brilliance calibration, the agent selects a standard Python least-squares solver. It also uses finite differences to exact the Jacobian of the model. Convergence of the fitting is evaluated through objective changes, scaled parameter steps, and a normalized gradient measure. For numerical validation, the agent reloads the exported lattice, reduces the finite-difference step, and compares predicted responses with independent small-kick calculations. The fitted optics and residuals are then compared with those from the specialist.

\bmhead{Acknowledgements}
This work is supported by the High Energy Photon Source (HEPS), a major national science and technology infrastructure. Y. Jiao's work is supported by the National Natural Science Foundation of China (No. 12275284). J. Wan's work is supported by the Talent Introduction Program of the IHEP (No. E6515BU1).

\section*{Declarations}

\subsection*{Funding}
This work was supported by the National Natural Science Foundation of China (No. 12275284) and the Talent Introduction Program of the IHEP (No. E6515BU1).

\subsection*{Conflict of interest/Competing interests}
The authors declare no competing interests.

\subsection*{Ethics approval and consent to participate}
Not applicable.

\subsection*{Consent for publication}
Not applicable.

\subsection*{Data availability}
The data that support the findings are all available at \cite{jinyu2026example}.

\subsection*{Materials availability}
Not applicable.

\subsection*{Code availability}
The Brilliance code is openly available on GitHub at \url{https://github.com/Jinyu95/Brilliance} under the MIT License.

\subsection*{Author contribution}
J. Wan conceived the study. J. Wan, W. Bao, H. Zhong and X. Lu implemented the Brilliance agent and performed the simulations. Y. Li performed the LOCO reference calibration. Y. Zhao analyzed the light source results. Y. Jiao supervised the study. J. Wan wrote the manuscript with input from all authors. All authors reviewed and approved the final manuscript.


\bibliography{sn-bibliography}

\end{document}